\documentclass[amsmath,amssymb,nofootinbib,twocolumn]{revtex4}
\usepackage{pgfplots}
\pgfplotsset{compat=1.17}
\usepackage{dcolumn}
\usepackage{color}
\usepackage{scrextend}
\usepackage{subfig}
\usepackage{bm}
\usepackage{amsmath,amssymb,graphicx,hyperref}
\usepackage{epsfig}
\usepackage{enumerate}
\usepackage{array}
\usepackage{multirow}
\usepackage{tikz}
\usepackage{ragged2e}
\usepackage{textgreek}
\usepackage[normalem]{ulem}

\begin{document}
\title
{Atomic-superfluid heat engines controlled by twisted light}
\author{Aritra Ghosh\footnote{aritraghosh500@gmail.com}, Nilamoni Daloi, and M. Bhattacharya}
\affiliation{School of Physics and Astronomy, Rochester Institute of Technology, 84 Lomb Memorial Drive, Rochester, New York 14623, USA}
\vskip-2.8cm
\date{\today}
\vskip-0.9cm

\vspace{5mm}
\begin{abstract}
We theoretically propose a quantum heat engine using a setup consisting of a ring-trapped Bose-Einstein condensate placed in a Fabry-P\'erot cavity where the optical field carries orbital angular momentum. We first show that the cavity-enhanced light-atom coupling leads to the emergence of polaritonic modes whose character can be reversibly switched between photonlike and phononlike by detuning sweeps, allowing work extraction governed by distinct reservoirs. We investigate the dependence of the engine efficiency on the orbital angular momentum. Beyond ideality, we discuss finite-time scenarios based on shortcuts to adiabaticity such that the efficiency retains its ideal-operation value, despite finite-time operation. Our analysis identifies orbital angular momentum as a control knob that can reconfigure the performance of such quantum heat engines.
\end{abstract}

\maketitle

\section{Introduction}
Quantum heat engines (QHEs) offer a tantalizing framework to explore the interplay between thermodynamics, quantum physics, and resource conversion in few-body systems \cite{Kosloff_2014,Cangemi_2024}. The canonical three-level-maser model investigated by Scovil and Schulz-DuBois \cite{Scovil_1959} already contained the essence of a QHE, in particular, the notions of energy quantization and the extraction of work from a quantum system. Subsequent formulations using the theory of open quantum systems \cite{Alicki_1979} have clarified the microscopic foundations of quantum cycles (Otto, Carnot, and Stirling), enabling performance analyses in the slowly-varying regime \cite{Cavina_2017,Scandi_2019} and beyond \cite{Feldmann_2000,Rezek_2006,Chen_2010,delCampo_2013,delCampo_2014}. Moreover, the ability to engineer non-thermal baths and to harness quantum correlations and coherence has revealed new bounds on the performance of mesoscopic and nanoscale devices \cite{Quan_2007,Scully_2001,Kosloff_2013,Guarnieri_2019,Bedkihal_2025}. Experimental progress in atomic and molecular systems has elevated QHEs from theoretical constructs to operational devices. Trapped-ion and single-atom engines have demonstrated full thermodynamic cycles \cite{Abah_2012,Rossnagel_2016,Maslennikov_2019}, verifying theoretical predictions. 

\vspace{2mm}

The versatility of atomic, molecular, and optical (AMO) platforms allows the implementation of Otto cycles with variable trap frequencies \cite{Kosloff_Rezek_2017}, optomechanical heat engines \cite{Zhang_2014}, quantum absorption refrigerators \cite{Maslennikov_2019,Ivander_2022}, and quantum devices under dynamical control \cite{Gelbwaser_2015}. These provide quantitative access to microscopic work statistics and coherence-assisted performance enhancements, thereby positioning atomic and molecular QHEs at the forefront of efforts to formulate a consistent formalism with predictive experimental relevance. In this direction, the use of ultracold atoms \cite{Brantut_2013,Barontini_2019} and Bose-Einstein condensates (BECs) \cite{Koch_2023,Simmons_2023,Ruan_2024} has found significant relevance as systems readily manipulatable in experimental settings, offering a rich platform for quantum phenomena.

\vspace{2mm}

\begin{figure}
\centering
\includegraphics[width=0.8\linewidth]{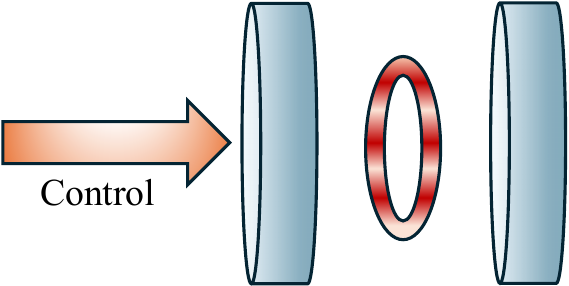}
\caption{\justifying{Schematic of the proposed setup. A BEC confined in a ring trap and rotating with winding number $L_p$ is placed inside a Fabry-P\'erot cavity. The cavity is driven by a two-tone control laser where each tone is prepared in coherent superpositions of Laguerre-Gaussian modes carrying OAM $\pm\ell\hbar$, generating a weak optical lattice that Bragg-diffracts atoms from the macroscopically-occupied persistent-current mode into two weakly-occupied sidemodes. Fluctuations of these sidemodes hybridize with the photonic fluctuations to form polaritonic excitations. By sweeping the control detuning, the photonlike and phononlike admixture of a polariton is reversibly switched, enabling the operation of a quantum Otto cycle.}}
\label{schematic}
\end{figure}

The aim of this work is to theoretically propose quantum heat engines based on the Otto cycle for toroidally-trapped \cite{Morizot_2006,Wright_2013} BECs that can be manipulated by orbital-angular-momentum (OAM)-carrying photons \cite{Molina-Terriza_2001,Yao_2011,Fickler_2012} in a Fabry-P\'erot cavity; see Fig. (\ref{schematic}). In this proposed setup, a BEC with a well-defined winding number/angular momentum is placed within a high-finesse cavity \cite{Brennecke_2008,Pandey_2019} so as to enhance its coupling to the intracavity photons. Such a setup has a strong potential for experimental realization \cite{Brennecke_2008,Pandey_2019} (see also, \cite{Fickler_2012}) and has been theoretically proposed to address several outstanding problems in AMO physics \cite{Kumar_2021,Kalita_2023,1_Pradhan_2024,2_Pradhan_2024,Das_2024,Gupta_2024,Pradhan_2025}. The cavity is driven by a two-tone control laser where each tone is prepared in a coherent superposition of Laguerre-Gaussian modes with OAM $\pm\ell\hbar$, thereby forming a weak sinusoidal optical potential around the ring, with the `weakness' ensured by blue-detuning the optical field far from the atomic resonance. This optical lattice acts as a Bragg grating, diffracting atoms out of the macroscopically-occupied persistent-current mode (winding number $L_p$) into two weakly-populated atomic sidemodes with winding numbers $L_p\pm 2\ell$ \cite{Kumar_2021}. Because these excitation processes must satisfy both angular momentum and energy conservation, the excitation of the sidemodes generally requires a two-photon process in which the optical fields compensate the atomic recoil energy. In the resolved-sideband (good-cavity) regime where the recoil energies exceed the cavity linewidth, this motivates the use of a two-tone control laser rather than a monochromatic one. Fluctuations of these sidemodes function as excitations that hybridize with the cavity-photon fluctuations to give rise to polariton normal modes. By performing detuning sweeps via external control, it is possible to perform a cyclic operation that performs work by operating between the phonon (hot) and photon (cold) reservoirs, much like a conventional heat engine. This net work obtained over a cyclic operation manifests as the energy delivered to the external device responsible for imposing the detuning sweeps. Being manipulated by cavity optical fields carrying OAM, the proposed quantum heat engine's efficiency shall be shown to be controllable by the OAM values of the optical fields, thereby opening up the possibility of new state-of-the-art quantum machines. 

\vspace{2mm}

The remainder of the paper is organized as follows. In Sec. (\ref{model_sec}), we shall briefly discuss the theoretical model on which our analysis is based. Then, in Sec. (\ref{polariton_sec}), we will describe the polariton modes arising due to the cavity-induced light-matter coupling and the framework of quantum Langevin equations, the latter leading naturally to thermodynamic notions. Sec. (\ref{otto_sec}) shall then be devoted to the description and analysis of an ideal quantum Otto cycle, wherein its efficiency along a single polariton branch will be quantified analytically. Finite-time (non-ideal) situations will be discussed in Sec. (\ref{finite_sec}). The paper is concluded in Sec. (\ref{conc_sec}) and the supplementary material is relegated to the Appendices (\ref{appA}), (\ref{appB}), (\ref{appC}), and (\ref{appD}).

\section{Theoretical model}\label{model_sec}
We will consider a BEC of $N$ identical $^{23}\mathrm{Na}$ atoms of mass $m$, confined in an annular ring trap \cite{Morizot_2006,Wright_2013} of radius $R$ and potential  $V(\rho) = \frac{1}{2}m\omega_\rho^2 (\rho-R)^2$, placed inside a Fabry-P\'erot cavity of length $L$, resonance frequency $\omega_0$, and photon decay rate $\gamma_0$, as illustrated in Fig. (\ref{schematic}). The cavity is driven through one of the mirrors by two coherent control tones with frequencies $\omega_{L1}$ and $\omega_{L2}$ and coherent drive strengths $\varepsilon_1$ and $\varepsilon_2$, respectively. Each control tone is prepared in a coherent superposition of Laguerre-Gaussian modes carrying OAM $\pm\ell\hbar$ \cite{Molina-Terriza_2001,Yao_2011,Fickler_2012}, generating a circular optical lattice about the cavity axis overlapping with the ring-shaped BEC.

\vspace{2mm}

The atoms undergo quantized rotational motion around the cavity axis, characterized by a winding number $L_p\in\mathbb{Z}$ \cite{Wright_2013}. The associated rotational energy is $\hbar\Omega_p$, where $\Omega_p=\hbar L_p^2/(2mR^2)$ \cite{Kumar_2021}. The Hamiltonian is $\hat{H} = \hat{H}_{\rm cavity} + \hat{H}_{\rm ring}$, with
\begin{equation}
   \frac{\hat{H}_{\rm cavity}}{\hbar} = -\Delta_{2}\hat{a}^{\dagger}\hat{a} +i\left(\varepsilon_{2} \hat{a}^\dagger - \varepsilon_{2}^* \hat{a}\right)
+i\left(\varepsilon_{1} e^{+i\delta t} \hat{a}^\dagger - \varepsilon_{1}^* e^{-i\delta t} \hat{a}\right),
\end{equation} expressed in the rotating frame of the second control tone, where $\Delta_{2}=\omega_{L2}-\omega_{0}$ is the detuning of the second tone from the cavity resonance and $\delta=\omega_{L2}-\omega_{L1}$ is the intertone spacing. The atomic Hamiltonian on the ring, taking into account two-body interactions, takes the generic form
\begin{eqnarray}
\hat{H}_{\rm ring} &=& \int_{0}^{2\pi} d\phi \hat{\Psi}^{\dagger}(\phi)\hat{\mathcal{H}}\hat{\Psi}(\phi) \nonumber\\
&&+ \frac{g}{2} \int_{0}^{2\pi} d\phi \hat{\Psi}^{\dagger}(\phi)\hat{\Psi}^{\dagger}(\phi)\hat{\Psi}(\phi)\hat{\Psi}(\phi),
\end{eqnarray}
where $\hat{\Psi}(\phi)$ satisfies $[\hat{\Psi}(\phi),\hat{\Psi}^{\dagger}(\phi')]=\delta(\phi-\phi')$ and $g=2\hbar\omega_{\rho}a_{\mathrm{Na}}/R$ is the effective interaction strength, with $a_{\mathrm{Na}}$ being sodium's $s$-wave scattering length. Employing a two-level approximation with dispersive light-matter coupling, the single-particle Hamiltonian density for angular motion along the ring on which the input light imprints an optical lattice is given by \cite{Kumar_2021,Gerry_Knight_2004}
\begin{eqnarray}
\hat{\mathcal{H}} &=&
-\frac{\hbar^{2}}{2mR^2}\frac{\partial^{2}}{\partial\phi^{2}}
+\hbar U_{0}\cos^{2}(\ell\phi)\hat{a}^{\dagger}\hat{a} ,
\label{Single_Particle_Hamiltonian}
\end{eqnarray}
where $U_0=g_a^2/\Delta_a$ denotes the single-photon light shift, with $g_a$ being the single-atom coupling and $\Delta_a$ being the atomic detuning. It is noteworthy that the control tones are classical coherent drives tuned near the same cavity resonance at frequency $\omega_0$ and populate a single quantized intracavity mode \cite{Aspelmeyer_2014} described by the operators $\hat{a}$ and $\hat{a}^\dagger$, satisfying $[\hat{a},\hat{a}^\dagger]=1$; in the rotating frame of one tone, the other one appears as a drive oscillating at the beat frequency $\delta=\omega_{L2}-\omega_{L1}$. 

\vspace{2mm}

The optical lattice formed by the control field causes Bragg scattering of atoms from their initial rotational mode with winding number $L_{p}$ to sidemodes with winding numbers $L_p\pm 2n\ell$ ($n=1,2,\cdots$). The optical fields are in the far-blue-detuned regime with respect to the atomic resonance making these diffractive effects weak so that it suffices to keep only $n=1$, leading to the ansatz
\begin{equation}
\hat{\Psi}(\phi)=\frac{1}{\sqrt{2\pi}}
\Big[e^{iL_{p}\phi}\hat{c}_{p}+e^{i(L_{p}+2\ell)\phi}\hat{c}_{+}+e^{i(L_{p}-2\ell)\phi}\hat{c}_{-}\Big],
\label{Ansatz}
\end{equation}
where $[\hat{c}_i,\hat{c}_j^\dagger]=\delta_{ij}$ for $i,j\in\{p,+,-\}$ and the expectation value of $\hat{c}_p^\dagger \hat{c}_p+\hat{c}_+^\dagger \hat{c}_+ + \hat{c}_-^\dagger \hat{c}_-$ is $N$. Since the original persistent-current mode is macroscopically occupied, one can treat it classically by imposing $\langle \hat{c}_p^\dagger \hat{c}_p\rangle \simeq N$, and introduce the sidemode operators $\hat{c}=\hat{c}_p^\dagger \hat{c}_+/\sqrt{N}$ and $\hat{d}=\hat{c}_p^\dagger \hat{c}_-/\sqrt{N}$, satisfying $[\hat{c},\hat{c}^\dagger]=[\hat{d},\hat{d}^\dagger]=1$ for large $N$. Under this approximation, the total Hamiltonian describing the optical mode and the two atomic sidemodes reads \cite{Kumar_2021}
\begin{eqnarray}
    \label{ring_BEC_Hamiltonian}
\frac{\hat{H}}{\hbar} &=&
- \tilde{\Delta}_2\hat{a}^\dagger \hat{a}
+ \omega_c \hat{c}^\dagger \hat{c}
+ \omega_d \hat{d}^\dagger \hat{d}
+ G(\hat{X}_c + \hat{X}_d)\hat{a}^\dagger \hat{a} \nonumber \\
&&\quad + i(\varepsilon_{2}\hat{a}^\dagger - \varepsilon^*_{2} \hat{a})
+ i(\varepsilon_{1}e^{+i\delta t}\hat{a}^\dagger - \varepsilon^*_{1}e^{-i\delta t} \hat{a}) \nonumber\\
&&\quad + 4\tilde{g}N(\hat{c}^\dagger \hat{c} + \hat{d}^\dagger \hat{d})
+ 2\tilde{g}N(\hat{c} \hat{d} + \hat{c}^\dagger \hat{d}^\dagger),
\end{eqnarray}
where $\hat{X}_{c(d)}=(\hat{c}_{(d)}+\hat{c}_{(d)}^\dagger)/\sqrt{2}$ are the sidemode position quadratures and $\omega_{c(d)}=\hbar[L_p\pm2\ell]^2/(2mR^2)$ are the corresponding sidemode frequencies. The light-matter coupling constant is $G=U_0\sqrt{N/8}$ \cite{Kumar_2021,Aspelmeyer_2014} and $\tilde{g}=g/(4\pi\hbar)$ denotes the strength of interatomic interactions. The effective detuning including the mean dispersive shift is $\tilde{\Delta}_2=\Delta_2-U_0N/2$. Across the parameter values considered here, the interaction-induced corrections associated with the last line of Eq. (\ref{ring_BEC_Hamiltonian}) are negligible (see Appendix (\ref{appA})), allowing us to drop them for our purposes.

\vspace{2mm}

With these simplifications, Eq. (\ref{ring_BEC_Hamiltonian}) reduces to a canonical optomechanical Hamiltonian with two mechanical modes. Let us linearize about the classical amplitudes $(\bar{a},\alpha_c,\alpha_d)$ by writing $\hat{a}=\bar{a}+\hat{\tilde{a}}$, $\hat{c}=\alpha_c+\hat{\tilde{c}}$, and $\hat{d}=\alpha_d+\hat{\tilde{d}}$, where the mean intracavity field contains both coherent components expressed as $\bar{a}(t)=\alpha_2+\alpha_1 e^{i\delta t}$. Retaining only bilinear terms in the fluctuation operators and working in the resolved-sideband regime ($\omega_{c,d}\gg\gamma_0$), one obtains after invoking the rotating-wave approximation, the following Hamiltonian (see Appendix (\ref{appB}) for details):
\begin{eqnarray}
\frac{\hat{H}}{\hbar}
&=&
-\bar{\Delta} \hat{a}^\dagger \hat{a}
+ \omega_c \hat{c}^\dagger \hat{c}
+ \omega_d \hat{d}^\dagger \hat{d} \nonumber\\
&& + \tilde{G}(\hat{a}^\dagger \hat{c} + \hat{a} \hat{c}^\dagger)
+ \tilde{G}(\hat{a}^\dagger \hat{d} + \hat{a} \hat{d}^\dagger),
\label{Hfull}
\end{eqnarray}
where we have simplified the notation by labeling $(\hat{\tilde{a}},\hat{\tilde{c}},\hat{\tilde{d}})=(\hat{a},\hat{c},\hat{d})$. Here, $\tilde{G}$ is the effective coupling constant defined in Appendix (\ref{appB}) and $\bar{\Delta}$ is the effective detuning including the (small) radiation-pressure-induced shift. The form of the Hamiltonian mentioned above shall form the basis of our analysis. In our analysis, we shall choose $m = 23$ amu, $R = 10~\mu m$, $L_p = 20$, $\ell \sim \mathcal{O}(10^2)$, and $\gamma_0 = 2\pi \times 10^3$ s$^{-1}$ \cite{Ludlow_2007}, making $\omega_c,\omega_d \gg \gamma_0$, strongly enforcing the resolved-sideband regime which justifies the rotating-wave approximation leading to the beam-splitter form of the Hamiltonian quoted above.

\section{Polariton modes}\label{polariton_sec}
\begin{figure}
    \centering
    \includegraphics[width=1\linewidth]{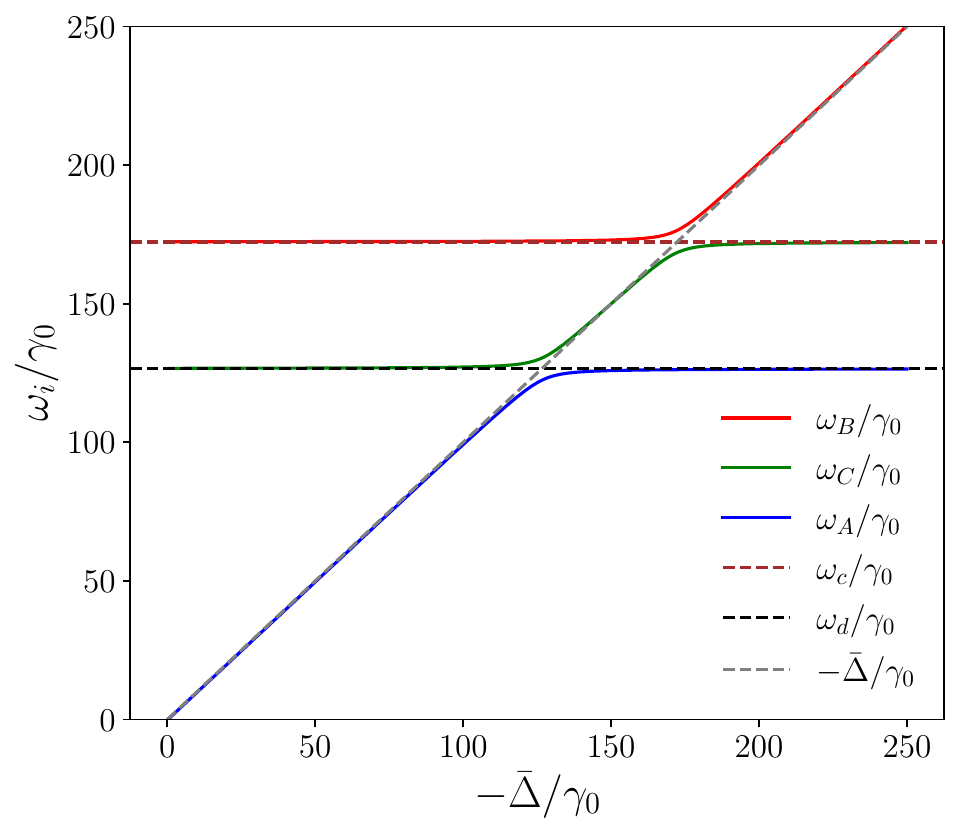}
    \caption{\justifying{Polaritonic frequencies (in units of $\gamma_0$) with $\tilde{G} = 4\gamma_0$, along with the bare modes $(\tilde{G}=0)$ for physical choices of the parameters conforming to $L_p = 20$, $\ell = 130$, $m = 23$ amu, and $R = 10~ \mu m$. This leads to $\omega_c \approx 173.27\gamma_0$ and $\omega_d \approx 126.56 \gamma_0$.}}
    \label{normal_final}
\end{figure}

With the physical picture and the identifications alluded to above, one can now describe the polariton modes formed due to the presence of the light-matter coupling. Considering the form of the Hamiltonian given in Eq. (\ref{Hfull}), the coefficient matrix  
\begin{equation} \Lambda = \begin{pmatrix} - \bar{\Delta} &  \tilde{G} &  \tilde{G} \\ \tilde{G} & \omega_c & 0 \\ \tilde{G} & 0 & \omega_d \end{pmatrix} ,\label{Lambda} \end{equation}
leads to the characteristic equation $(-\bar{\Delta} - \lambda)(\omega_c - \lambda)(\omega_d - \lambda)
- \tilde{G}^2\big[(\omega_c - \lambda) + (\omega_d - \lambda)\big] = 0$, giving rise to three real and generically non-degenerate eigenvalues corresponding to the three polariton modes (see Appendix (\ref{appC})). These normal-mode frequencies have been plotted in Fig. (\ref{normal_final}) as a function of $-\bar{\Delta}$, all in units of the photon damping rate $\gamma_0$. 
We will label the lower mode `$A$' with frequency $\lambda_A = \omega_A$, middle mode `$C$' with frequency $\lambda_C =\omega_C$, and upper mode `$B$' with frequency $\lambda_B =\omega_B$. In the limit $\tilde{G}\to 0$, one gets the limiting values, $\omega_A\to -\bar{\Delta}$, $\omega_B\to\omega_c$, and $\omega_C\to\omega_d$, i.e., the bare modes, also shown in Fig. (\ref{normal_final}) with dashed lines. By a direct inspection of Fig. (\ref{normal_final}), clearly, the `$A$' mode is photonlike $\sim \mathcal{O}(-\bar{\Delta})$  for small $-\bar{\Delta} >0$ but behaves in a phononlike manner $\sim \mathcal{O}(\omega_d)$ for large $-\bar{\Delta} > 0$. The opposite is true for the mode `$B$'. The reader is referred to Appendix (\ref{appD}) for the asymptotic expressions of the polaritonic frequencies in the limits $|\bar{\Delta}| \ll \omega_{c,d}$ and $|\bar{\Delta}| \gg \omega_{c,d}$, clearly revealing the above-mentioned asymptotic regimes.

\subsection{Normal-mode transformations}
The normal-mode transformations which diagonalize the linearized Hamiltonian in Eq. (\ref{Hfull}) are
\begin{eqnarray}
\hat{A}&=&\frac{1}{N_A}\left(\hat{a}-\frac{\tilde{G}}{\omega_c-\omega_A} \hat{c}-\frac{\tilde{G}}{\omega_d-\omega_A} \hat{d}\right), \label{bogA}\\
 \hat{B}&=&\frac{1}{N_B}\left(\hat{a}-\frac{\tilde{G}}{\omega_c-\omega_B} \hat{c}-\frac{\tilde{G}}{\omega_d-\omega_B} \hat{d}\right),\\
  \hat{C}&=&\frac{1}{N_C}\left(\hat{a}-\frac{\tilde{G}}{\omega_c-\omega_C} \hat{c}-\frac{\tilde{G}}{\omega_d-\omega_C} \hat{d}\right),
\end{eqnarray}
with $N_j \equiv N(\omega_j) = \sqrt{1+\frac{\tilde{G}^2}{(\omega_c-\omega_j)^2}+\frac{\tilde{G}^2}{(\omega_d-\omega_j)^2}}$, for $j = A,B,C$. It is straightforward to verify the unitary (number-conserving) nature of these transformations as $\hat{A}^\dagger \hat{A} + \hat{B}^\dagger \hat{B} + \hat{C}^\dagger \hat{C} = \hat{a}^\dagger \hat{a} + \hat{c}^\dagger \hat{c} + \hat{d}^\dagger \hat{d}$. Equivalently, the thermal expectation values satisfy
\begin{equation}
\langle   \hat{A}^\dagger  \hat{A}\rangle+\langle   \hat{B}^\dagger  \hat{B}\rangle + \langle \hat{C}^\dagger  \hat{C}\rangle
= n_a+n_c+n_d,
\end{equation}
where $n_{a,c,d}$ are the Bose factors corresponding to the bare modes $a$, $c$, and $d$. For a given polaritonic frequency $j$, the Hopfield coefficients \cite{Hopfield_1958,Kavokin_Malpuech_2003} are easily found to be
\begin{eqnarray}
X_a^{(j)}&=&\frac{1}{N(\omega_j)},\\
X_c^{(j)}&=&\frac{\tilde{G}}{(\omega_j-\omega_c)N(\omega_j)},\\
X_d^{(j)}&=&\frac{\tilde{G}}{(\omega_j-\omega_d)N(\omega_j)},
\label{Hopfield_eqn}
\end{eqnarray} whence, it follows that $|X_a^{(j)}|^2+|X_c^{(j)}|^2+|X_d^{(j)}|^2=1$, as required by construction. These coefficients exactly determine the polaritonic excitations as coherent superpositions of the excitations of the photon mode and the two atomic sidemodes. This decomposition transparently reveals the coupling of a polariton to the photon and phonon reservoirs and its ability to switch between photonlike and phononlike character under detuning sweeps.

\vspace{2mm}

Let us restrict our focus to the lower polariton branch, i.e., the mode $A$. Since $\tilde{G} \ll \omega_{c,d}$, one can perturbatively determine the behavior of $\omega_A$ for large and small values of the detuning (see Appendix (\ref{appD})). It can be shown that when $-\bar{\Delta}  \ll \omega_{d}$, we have
\begin{equation}\label{omegaAsmall}
    \omega_A \simeq -\bar{\Delta}-\frac{\tilde{G}^2}{\bar{\Delta} +\omega_c}-\frac{\tilde{G}^2}{\bar{\Delta} +\omega_d},
\end{equation}
from where its photonlike nature is transparent, along with the AC-Stark shift due to light-matter coupling. On the other extreme, i.e., for $-\bar{\Delta}  \gg \omega_{d}$, one has
\begin{equation}\label{omegaAlarge}
    \omega_A \simeq \omega_d+\frac{\tilde{G}^2}{\bar{\Delta}+\omega_d},
\end{equation}
whence its phononlike ($d$-like) nature is analytically revealed, accompanied by the AC-Stark shift. Thus, following the proposal of \cite{Zhang_2014}, we can now describe a quantum heat engine operating along the polariton branch $A$. Prior to that, however, we shall first describe some generic quantum-thermodynamic features relying on the stochastic framework \cite{{Gardiner_Zoller_2004},Seifert_2012} that incorporates the fluctuation-dissipation theorem. 

\subsection{Quantum Langevin equations and stochastic considerations}
 Since the system is not completely isolated and is influenced by dissipative effects, it is important to take into account the effects due to the environment in terms of the photon and phonon reservoirs. The quantum Langevin equations, along with the built-in fluctuation-dissipation theorem, can describe these environmental effects. Considering the bare modes, these are given by \cite{Gardiner_Zoller_2004}
 \begin{eqnarray}
  \frac{d\hat{a}}{dt} - i \bar{\Delta} \hat{a} + \bigg(\frac{\gamma_0}{2}\bigg) \hat{a} = \sqrt{\gamma_0} \hat{a}_{\rm in}(t), \label{qle1}\\
    \frac{d\hat{c}}{dt} + i \omega_c  \hat{c} + \bigg(\frac{\gamma_m}{2}\bigg) \hat{c} = \sqrt{\gamma_m} \hat{c}_{\rm in}(t), \label{qle2} \\
      \frac{d\hat{d}}{dt} + i \omega_d \hat{d} + \bigg(\frac{\gamma_m}{2}\bigg) \hat{d} = \sqrt{\gamma_m} \hat{d}_{\rm in}(t), \label{qle3}
 \end{eqnarray} where $\gamma_0$ and $\gamma_m$ are the photon and phonon damping rates, while $\{\hat{a}_{\rm in}(t), \hat{c}_{\rm in}(t), \hat{d}_{\rm in}(t)\}$ are the input noises \cite{Aspelmeyer_2014}. For the cavity photons, we shall take $\gamma_0 = 2\pi$ kHz \cite{Ludlow_2007}. In contrast, the damping rate of the atomic sidemodes, which we shall take to be given by $\gamma_m = 1.7 \times 10^{-5} \gamma_0$, is set by the lifetime of persistent currents \cite{Moulder_2012}, leading to a slower equilibration with the phonon bath. The fluctuations captured by the noisy effects are related to the damping (dissipative) constants via the fluctuation-dissipation relations provided by
 \begin{eqnarray}
 \langle \hat{a}_{\rm in}^\dagger(t) \hat{a}_{\rm in}(t') \rangle &=&  n_a \delta(t-t'), \\
  \langle \hat{c}_{\rm in}^\dagger(t) \hat{c}_{\rm in}(t') \rangle &=&  n_c \delta(t-t'), \\
   \langle \hat{d}_{\rm in}^\dagger(t) \hat{d}_{\rm in}(t') \rangle &=& n_d \delta(t-t'),   
 \end{eqnarray} where $n(\omega,T) = \big[\exp[{\hbar \omega/k_B T}] - 1\big]^{-1}$ is the Bose factor. The above relations ensure that the quantum Langevin equations [Eqs. (\ref{qle1}), (\ref{qle2}), and (\ref{qle3})] consistently give rise to $\langle \hat{a}^\dagger \hat{a} \rangle = n_a$, $\langle \hat{c}^\dagger \hat{c} \rangle = n_c$, and $\langle \hat{d}^\dagger \hat{d} \rangle = n_d$, where the angled brackets $\langle \cdot \rangle$ denote thermal averaging, i.e., averaging over the noise statistics stemming from arguments based on the ergodic hypothesis. For optical frequencies, the photon occupation number can be taken to be zero \cite{Aspelmeyer_2014}, indicating that $T_{\rm photon} \approx 0$ K. The temperature of the phonon bath is, however, non-zero, being of the order of $10^2$ nK \cite{Pethick_Smith_2001}. 
 
 \vspace{2mm}
 
 Let us now demonstrate how the above-mentioned fluctuations in the bare modes impact those in the polaritonic modes, focusing on the lower polariton branch. The fluctuating and dissipative effects can be phenomenologically captured by the following Langevin-type equation:
 \begin{equation}
   \frac{d\hat{A}}{dt} + i \omega_A \hat{A} + \bigg(\frac{\gamma_{\rm eff}}{2}\bigg) \hat{A} = \sqrt{\gamma_{\rm eff}} \hat{A}_{\rm in}(t), \label{qle4}
 \end{equation}
 where $\gamma_{\rm eff}$ is the effective damping constant and $\hat{A}_{\rm in}(t)$ is an effective input noise that includes the photonlike and phononlike fluctuations, mixed appropriately by the Hopfield weights. It is reasonable to assume that the effective input noise is delta-correlated, i.e., its power spectrum is a constant. The above-mentioned equation can be solved easily by going into the Fourier domain and the quantity $\langle \hat{A}^\dagger(t) \hat{A}(t') \rangle$ for $t=t'$ can be found by invoking the Wiener-Khinchin theorem. The result implies that the correlation function of the input noise is given by
 \begin{equation}
 \langle \hat{A}^\dagger_{\rm in}(t) \hat{A}_{\rm in}(t') \rangle = \langle \hat{A}^\dagger \hat{A} \rangle \delta(t-t'),
 \end{equation} where $\langle \hat{A}^\dagger \hat{A} \rangle$ is determined by the transformation given in Eq. (\ref{bogA}) which gives
 \begin{equation}\label{AdaggerA}
\langle \hat{A}^\dagger \hat{A} \rangle =  \left[ \frac{n_a+\dfrac{\tilde{G}^2}{(\omega_c-\omega_A)^2}n_c+\dfrac{\tilde{G}^2}{(\omega_d-\omega_A)^2}n_d}
{1+\dfrac{\tilde{G}^2}{(\omega_c-\omega_A)^2}+\dfrac{\tilde{G}^2}{(\omega_d-\omega_A)^2}} \right],
 \end{equation} playing the role of a fluctuation-dissipation theorem; we can further set $n_a=0$ as discussed earlier. In obtaining the above-mentioned expression, we have assumed that the cross-correlations between the bare modes vanish. This is justified because (a) we have assumed independent photon and phonon reservoirs so that $\langle \hat{a}^\dagger_{\rm in}(t)\hat{c}_{\rm in}(t')\rangle=\langle \hat{a}^\dagger_{\rm in}(t)\hat{d}_{\rm in}(t')\rangle=0$, (b) in the Born-Markov-secular limit and for $|\omega_c-\omega_d|$ very large compared to the inverse bath-correlation rate (which is indeed the case here), the nonsecular terms average out, yielding $\langle \hat{c}^\dagger_{\rm in}(t)\hat{d}_{\rm in}(t')\rangle = \langle \hat{d}^\dagger_{\rm in}(t)\hat{c}_{\rm in}(t')\rangle=0$. Intuitively, orthogonal angular-momentum modes see uncorrelated local noise, and their widely-separated frequencies ($|\omega_c - \omega_d| \gg \gamma_m$) make cross-correlation terms secularly negligible; see Chapter 3 of \cite{Breuer_Petruccione_2002} for more details. In other words, for \(\omega_{c,d} \sim \mathcal{O}(10^2 \gamma_0)\), the dimensionless parameter 
 $\frac{\gamma_m}{|\omega_c - \omega_d|}$ that signifies the strength of cross-correlations is tiny, i.e., $\frac{\gamma_m}{|\omega_c - \omega_d|} \sim \frac{10^{-5} \gamma_0}{10^2 \gamma_0} \sim 10^{-7}$. Cross-correlations between the sidemodes may become significant only when $\omega_c \approx \omega_d$, which, however, never happens for our parameters.
 
 \vspace{2mm}
 
 Having now set up the quantum Langevin framework, let us consider the possibility of performing useful work by manipulating the lower polariton branch described by the Hamiltonian $\hat{H}_A = \hbar \omega_A \hat{A}^\dagger \hat{A}$. Considering a situation where $\omega_A$ is varied via variations of $\bar{\Delta}$, one can therefore write $d\hat{H}_A = \hbar d\omega_A (\hat{A}^\dagger \hat{A}) + \hbar \omega_A (\hat{A}^\dagger d\hat{A} + d\hat{A}^\dagger \hat{A})$, which, upon substituting the quantum Langevin equation [Eq. (\ref{qle4})] and supplemented by its Hermitian conjugate, simplifies to the intuitive form
 \begin{equation}
 \Delta \hat{\mathcal{Q}} = d\hat{H}_A + \Delta \hat{\mathcal{W}},
 \end{equation}
 where the incremental heat and work operators are identified to be $\Delta \hat{\mathcal{Q}} =  \hbar \omega_A \big[-\gamma_{\rm eff} \hat{A}^\dagger \hat{A} + \sqrt{\gamma_{\rm eff}} \hat{A}^\dagger  \hat{A}_{\rm in}(t) + \sqrt{\gamma_{\rm eff}} \hat{A}^\dagger_{\rm in}(t) \hat{A} \big] dt$ and $\Delta \hat{\mathcal{W}} = - \hbar d\omega_A (\hat{A}^\dagger \hat{A})$, respectively. The former involves both loss of energy into the environment as captured by $\gamma_{\rm eff}$ and the energy pumped into the system due to the input fluctuations. Notice that while representing energy balance, these operators incorporate stochastic fluctuations and thermodynamic interpretations can be assigned only after averaging over appropriate distributions. Averaging over the noise ensemble as indicated by $\langle \cdot \rangle$, one finds the first law of thermodynamics $\Delta Q = dE + \Delta W$; in particular, $E \equiv \langle \hat{H}_A \rangle$ and $\Delta W \equiv \langle \Delta \hat{\mathcal{W}}\rangle$, both of which now have thermodynamic meanings. While similar treatments can be made for the other two polaritonic modes, the above-mentioned setup suffices for the description of the quantum Otto cycle as shall be outlined in the following section. In particular, the hybrid nature of the polariton which switches between the phononlike and photonlike regimes via detuning sweeps suggests that it may be utilized as a working substance between two reservoirs, namely, the phonon and photon reservoirs. In an experiment, such sweeps of the control detuning can be implemented by adjusting either the cavity length or the drive-laser frequency, and the corresponding actuator constitutes the physical channel through which the engine's work output may be extracted.
 
\section{Ideal Otto cycle}\label{otto_sec}
We are now in the position to describe an ideal Otto cycle working along the lower polariton branch $A$. A schematic of the cycle is given below. 
    \noindent
\begin{align} \nonumber
&\begin{array}{cccccc}
\bar{\Delta}_i,T_i~&  &\text{Isentropic}& &&\bar{\Delta}_f,T_i\\
& (a) &\longrightarrow& (b)&&\\
&   &&\\
 \text{Isochoric}&\uparrow&&\downarrow&&\text{Isochoric}\\
&   &&\\
& (d)&\longleftarrow&(c) &&\\
\bar{\Delta}_i,T_f~& &\text{Isentropic}& &&\bar{\Delta}_f,T_f\\
&&&&&\\ \end{array} \\
&~~~~~~~~~-\bar{\Delta}_i  \gg \omega_{c,d};~~~ -\bar{\Delta}_f  \ll \omega_{c,d}
\nonumber
\end{align}
Under idealized and quasistatic conditions, each step can be understood as follows:

\begin{enumerate}
    \item \textbf{Isentropic expansion:} Starting with an initial value of the detuning compatible with $-\bar{\Delta}_i  \gg \omega_{c,d} $, i.e., in the regime where the polariton describes to an excellent approximation, phononlike excitations, the detuning is taken to a final value $-\bar{\Delta}_f  \ll \omega_{c,d}$. In other words, the detuning takes the polariton from the phononlike regime to the photonlike regime. Let $T_i$ be the initial temperature at the beginning of this step and $\Omega_{i}$ be the corresponding eigenfrequency. Ideally, the transformation is carried out adiabatically so that the polaritonic particle number $\langle \hat{A}^\dagger \hat{A} \rangle$ remains fixed and no heat is exchanged with external reservoirs. 

    \item \textbf{Isochoric transition:} In the photonlike regime, the system is allowed to relax with respect to the photonlike reservoir, i.e., at temperature $T_f \approx 0$ K. The thermal expectation value $\langle \hat{A}^\dagger \hat{A} \rangle$ adjusts to the new eigenfrequency $\Omega_{f}$ and temperature $T_f$ during this process. Ideally, full thermalization is desired. 

    \item \textbf{Isentropic compression:} From the detuning at $\bar{\Delta}_f$, it is taken back to its initial value $\bar{\Delta}_i$ (reverse of step 1) adiabatically so that $\langle \hat{A}^\dagger \hat{A} \rangle$ remains fixed and there is no heat exchange with the reservoirs within idealized conditions. 

    \item \textbf{Isochoric transition:} Back into the phononlike regime, the system is allowed to relax with respect to the phononlike reservoir, i.e., at temperature $T_i$. The thermal expectation value $\langle \hat{A}^\dagger \hat{A} \rangle$ adjusts to the eigenfrequency $\Omega_{i}$ and temperature $T_i$ upon full thermalization under ideal conditions. 
\end{enumerate}

\begin{figure}
\centering
\includegraphics[width=1\linewidth]{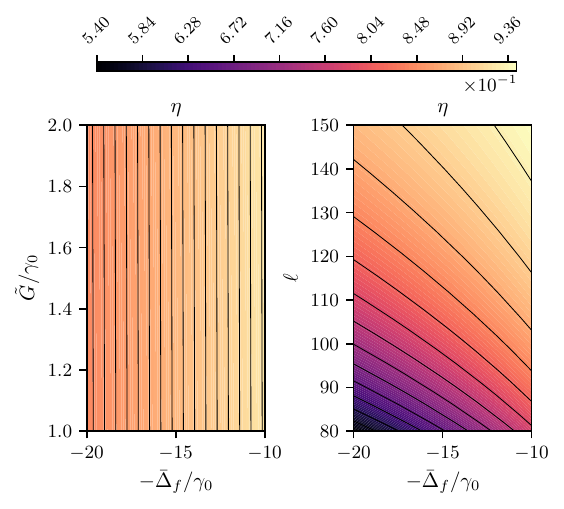}
\caption{\justifying{Variation of the efficiency of the Otto cycle as a function of the detuning $-\bar{\Delta}_f$ (in units of $\gamma_0$) along with the light-matter coupling constant $\tilde{G}$ (in units of $\gamma_0$; left panel) and the orbital angular momentum $\ell$ (right panel). We have taken $|\bar{\Delta}_i| = 2 \omega_d$. The left panel corresponds to $\ell = 130$ while the right panel corresponds to $\tilde{G} = 2\gamma_0$.}}
\label{eta_2D_plot}
\end{figure}

This cycle performs useful work due to variation of the detuning that completes the thermodynamic cycle. The thermal efficiency can be obtained by taking the ratio between the work performed by the engine in one cycle and the input heat. In order to determine the work done, let us resort to the prescription introduced earlier in which the incremental work done is expressible in the manner 
\begin{eqnarray}
    \Delta W = - \hbar \langle \hat{A}^\dagger \hat{A} \rangle d\omega_A. 
\end{eqnarray}
Clearly, the engine performs no work during the isochoric processes $(b) \rightarrow (c)$ and $(d) \rightarrow (a)$. So the total work done is found by
\begin{equation}
W = - \int_{(a)}^{(b)} \hbar \langle \hat{A}^\dagger \hat{A} \rangle d\omega_A  - \int_{(c)}^{(d)} \hbar \langle \hat{A}^\dagger \hat{A} \rangle d\omega_A. 
\end{equation}
Based on the earlier-stated assumption that these processes allow $\langle \hat{A}^\dagger \hat{A} \rangle$ to remain constant, also called the quantum adiabatic approximation, one gets
\begin{eqnarray}
W &\approx& - \langle \hat{A}^\dagger \hat{A} \rangle_i \int_{\Omega_i}^{\Omega_f} \hbar d\omega_A  - \langle \hat{A}^\dagger \hat{A} \rangle_f \int_{\Omega_f}^{\Omega_i} \hbar d\omega_A  \nonumber \\
&=& \hbar  (\Omega_i - \Omega_f) \big[\langle \hat{A}^\dagger \hat{A} \rangle_i - \langle \hat{A}^\dagger \hat{A} \rangle_f\big]. \label{workdefinitionconsistent}
\end{eqnarray}
\begin{widetext}
Referring to Eq. (\ref{AdaggerA}), therefore, the work done admits the following closed-form expression:
\begin{eqnarray}\label{work_otto_ana}
 \frac{W}{\hbar} = (\Omega_i - \Omega_f) \left[\frac{\dfrac{\tilde{G}^2}{(\omega_{c}-\Omega_i)^2}n_c+\dfrac{\tilde{G}^2}{(\omega_d-\Omega_i)^2}n_d}
{1+\dfrac{\tilde{G}^2}{(\omega_c-\Omega_i)^2}+\dfrac{\tilde{G}^2}{(\omega_d-\Omega_i)^2}} - \frac{\dfrac{\tilde{G}^2}{(\omega_c-\Omega_f)^2}n_c+\dfrac{\tilde{G}^2}{(\omega_d-\Omega_f)^2}n_d}
{1+\dfrac{\tilde{G}^2}{(\omega_c-\Omega_f)^2}+\dfrac{\tilde{G}^2}{(\omega_d-\Omega_f)^2}}\right],
\end{eqnarray}
where we have taken $n_a \approx 0$, suppressed by an astronomically-large factor because $T_{\rm photon} \approx 0$ K at optical frequencies.
\end{widetext}
Now, in order to quantify the heat exchanged with the environment, let us first note that during the processes $(a) \rightarrow (b)$ and $(c) \rightarrow (d)$, no heat is exchanged. For the remaining two, the heat exchanged is found readily by resorting to the first law of thermodynamics $Q = \Delta E$. Explicitly, one gets 
\begin{eqnarray}
Q_{\rm in} = - \hbar \Omega_i ( \langle \hat{A}^\dagger \hat{A} \rangle_f - \langle \hat{A}^\dagger \hat{A} \rangle_i ), \label{heatexpression} \\
Q_{\rm out} =  \hbar \Omega_f (\langle \hat{A}^\dagger \hat{A} \rangle_f - \langle \hat{A}^\dagger \hat{A} \rangle_i). \nonumber
\end{eqnarray}
Notice that since $\langle \hat{A}^\dagger \hat{A} \rangle_i > \langle \hat{A}^\dagger \hat{A} \rangle_f$, $Q_{\rm in} > 0$. Combining the expression for $Q_{\rm in}$ found above with Eq. (\ref{workdefinitionconsistent}) for the work done, the efficiency emerges to be
\begin{eqnarray}\label{otto_eff_general}
  \eta = \frac{W}{Q_{\rm in}} = 1 - \frac{\Omega_f}{\Omega_i} ,
\end{eqnarray}
consistent with earlier treatments of Otto cycles with quantum oscillators \cite{Kosloff_Rezek_2017}. The behavior of the efficiency has been depicted in Fig. (\ref{eta_2D_plot}) showing that theoretically, an excellent  efficiency can be achieved by appropriate tuning of the underlying parameters. Referring to the asymptotic expressions made explicit in Eqs. (\ref{omegaAsmall}) and (\ref{omegaAlarge}), a simple calculation reveals that the efficiency goes as
\begin{equation}
\eta \approx 1 + \frac{\bar{\Delta}_f}{\omega_d}
+ \frac{\tilde{G}^{2}}{\omega_d} \left(
\frac{1}{\omega_d}
+ \frac{1}{\omega_c}
\right),
\end{equation}
up to second order in $\tilde{G}$, upon using the facts that $\tilde{G} \ll \omega_{c,d}$, $-\bar{\Delta}_i  \gg \omega_{c,d} $, and $-\bar{\Delta}_f  \ll \omega_{c,d}$. One can note, in particular, that increased values of the OAM ($\ell$) gives rise to improved efficiencies, making it a tunable control parameter to boost the performance of such a device. 
Although this seemingly-innocuous framework described above can naively lead to efficiencies arbitrarily close to one, it should be noted that the other two polariton branches will also have to be taken into account. In particular, the branch $B$ which behaves in a complementary fashion compared to $A$ gives rise to negative values of the work done and this would decrease the overall efficiency of the cavity-controlled BEC heat engine. Thus our results as discussed so far neither violate the principles of thermodynamics, nor should they be interpreted as surpassing the known limits on the efficiency of quantum heat engines. The novel aspect, however, is the significant control on the performance of such an atomic quantum heat engine rendered by the OAM degree of freedom, easily controllable in an experimental setup involving ultracold atoms and an optical cavity. 

\section{Finite-time effects}\label{finite_sec}
In any practical implementation, the conditions are non-ideal and this leads to departures from theoretically-estimated efficiencies derived under the assumption of ideality. However, it is possible to analytically characterize the effect of certain non-ideal working conditions. In the present situation, there are mainly two sources of non-ideality, namely, (a) that the isentropic steps may not be completely adiabatic, and (b) that the isochoric steps require finite operating times for practical implementation. The second one is critical: full thermalization with the phonon bath would require an isochore over the timescale $\gamma_m^{-1}$, i.e., comparable to the persistent-current lifetime, which would compromise the condensate. Hence, realistic operation necessarily involves incomplete thermalization.

\subsection{Finite-time isentropes: Shortcuts to adiabaticity}
Let us begin by addressing the first source of imperfection. During the isentropic strokes, the frequency $\Omega$ is modulated between $\Omega_{\rm photon}$ and $\Omega_{\rm phonon}$, i.e., $\omega_A$ switches between the photonlike ($\sim -\bar{\Delta}$) and phononlike ($\sim \omega_d$) regimes. As assumed in Sec. (\ref{otto_sec}), in the quantum adiabatic limit, the polaritonic particle number in the instantaneous energy basis remains invariant and the energy scales linearly with $\Omega$. For finite-time driving, however, nonadiabatic effects are present \cite{Rezek_2006,Chen_2010,delCampo_2013,delCampo_2014}. The effect of these excitations can be compactly described by the adiabaticity parameter \cite{Husimi_1953} $Q^*\geq 1$, with $Q^* = 1$ for perfect adiabaticity. Nevertheless, the effect of such imperfections may be avoided by operating via adiabatic shortcuts \cite{Chen_2010,delCampo_2013,delCampo_2014,Kiran_2021}, effectively leading to $Q^* \approx 1$ even for finite-time operations. The essential idea is to consider a time-dependent oscillator (here, a polariton mode) $\ddot{\hat{X}}_A + \Omega(t)^2 \hat{X}_A = 0$, where $\Omega(t)$ is varied from $\Omega_i$ to $\Omega_f$, and $\hat{X}_A$ is the position operator of the mode. This description is robust if the timescale over which a detuning sweep is implemented is slow enough to suppress transitions between the polaritonic branches, separated near the avoided crossings by $\sim \mathcal{O}(\tilde{G})$. Based on the theory of the time-dependent quantum oscillator and the Ermakov-Lewis invariant, the expectation value of the Hamiltonian in the $n$th state reads \cite{Chen_2010}
\begin{equation}
    \overline{\hat{H}_A(t)}_n = \frac{(2n+1)\hbar}{4 \Omega_i} \bigg( \dot{\rho}(t)^2 + \Omega(t)^2 \rho(t)^2 + \frac{\Omega_i^2}{\rho(t)^2} \bigg),
\end{equation} where $\rho(t)$ is a scaling factor satisfying the Ermakov-Pinney equation \cite{Morris_2015}
\begin{equation}\label{ErmakovPinney}
    \ddot{\rho}(t) + \Omega(t)^2 \rho(t) - \frac{\Omega_i^2}{\rho(t)^3} = 0.
\end{equation}
Leaving the finite-time protocol $\Omega = \Omega(t)$ unspecified at the moment, one can impose boundary conditions on $\rho(t)$ and its time derivatives at the initial and final times $t_i = 0$ and $t_f = \tau$, respectively, to ensure that any eigenstate of $\hat{H}_A(0)$ evolves as a single expanding mode and it becomes (up to a phase factor) equal to the corresponding eigenstate of the $\hat{H}_A(\tau)$. This keeps the populations in the instantaneous basis equal at the initial and final times, exactly as one desires. The function $\rho(t)$ may be chosen as a real-valued function satisfying the boundary conditions set earlier and then the exact protocol $\Omega(t)$ can be determined from Eq. (\ref{ErmakovPinney}). 

\vspace{2mm}

\begin{figure}
    \centering
    \includegraphics[width=\linewidth]{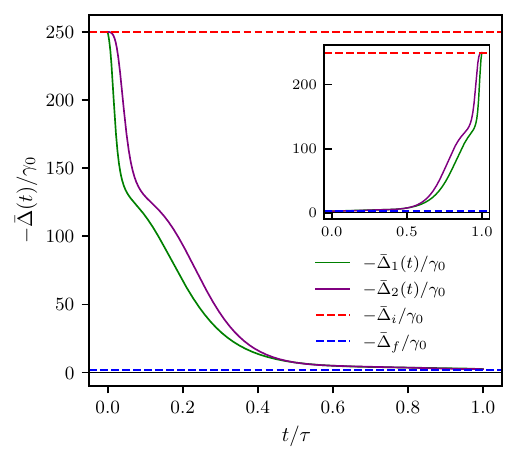}
    \caption{\justifying{Finite-time detuning protocols for the isentropes $(a) \rightarrow (b)$ (main figure) and $(c) \rightarrow (d)$ (inset). Here, $-\bar{\Delta}_{1}(t)$ and $-\bar{\Delta}_{2}(t)$ represent the detuning protocols corresponding to the frequency protocols $\Omega_1(t)$ and $\Omega_2(t)$ evaluated using $\rho_1(t)$ given in Eq. (\ref{rho_1}) and $\rho_2(t)$ given in Eq. (\ref{rho_2}), respectively. The dashed lines represent the detunings $|\bar{\Delta}_i| = 250\gamma_0$ and $|\bar{\Delta}_f| = 2\gamma_0$.}}
    \label{fig:placeholder}
\end{figure}

Taking $\rho(0) = 1$, $\dot{\rho}(0) = 0$, $\ddot{\rho}(0) = 0$, $\rho(\tau) = \sqrt{\Omega_i/\Omega_f}$, $\dot{\rho}(\tau) = 0$, and $\ddot{\rho}(\tau) = 0$, an explicit calculation taking a polynomial ansatz for $\rho(t)$ reveals its structure to be \cite{Chen_2010}
\begin{eqnarray}
\label{rho_1}
    \rho_1(t) &=& 6 \left( \sqrt{\frac{\Omega_i}{\Omega_f}} - 1 \right) \left(\frac{t}{\tau}\right)^5 - 15 \left( \sqrt{\frac{\Omega_i}{\Omega_f}} - 1 \right) \left(\frac{t}{\tau}\right)^4 \nonumber \\ 
    &&~ + 10 \left( \sqrt{\frac{\Omega_i}{\Omega_f}} - 1 \right) \left(\frac{t}{\tau}\right)^3 + 1.
\end{eqnarray}  While a polynomial form of $\rho(t)$ is commonly used in the literature \cite{Chen_2010,delCampo_2013,Kiran_2021}, it may be noted that the above-mentioned form is by no means unique; for instance, the following is also a plausible choice:
\begin{eqnarray}
\label{rho_2}
    \rho_2(t) &=& 1 + \left(\sqrt{\frac{\Omega_i}{\Omega_f}} - 1\right) \times \nonumber\\
   &&~ \left[ \frac{1}{2} - \frac{9}{16} \cos \left(\frac{\pi t}{\tau}\right) + \frac{1}{16} \cos \left( \frac{3 \pi t}{\tau} \right) \right],
\end{eqnarray}
satisfying the same boundary conditions. The variation $\Omega=\Omega(t)$ from the initial value to the final one is given by putting the function $\rho(t)$ into Eq. (\ref{ErmakovPinney}). The corresponding detuning protocol can be determined from the variation of $\Omega = \omega_A(-\bar{\Delta})$ as depicted in Fig. (\ref{normal_final}), blue curve. This has been shown in Fig. (\ref{fig:placeholder}) for the two protocols indicated above. 
\vspace{2mm}

\subsection{Finite-time isochores: Incomplete thermalization}
Let us now come to the isochoric strokes to analyze the finite-time effects. Ideally, the polaritonic mode should thermalize completely with respect to the photon/phonon reservoirs. In realistic operations, however, this is not true because these steps are operated over a finite window of time. Moreover, as mentioned earlier, since the phonon bath has a relaxation timescale of $\gamma_m^{-1}$, complete thermalization will require the step $(d) \rightarrow (a)$ to be performed over a timescale $\sim \gamma_m^{-1}$ to allow complete thermalization, but which will also degrade the condensate as the persistent currents have a lifetime of $\gamma_m^{-1}$. Taking into account the relaxation dynamics, the polaritonic particle number $\langle \hat{A}^\dagger \hat{A}\rangle$ undergoes thermalization over timescale $\tau$ as
\begin{equation}
    \langle \hat{A}^\dagger \hat{A} \rangle(\tau) = \langle \hat{A}^\dagger \hat{A} \rangle_{\rm final} + \left[ \langle \hat{A}^\dagger \hat{A} \rangle_{\rm initial} - \langle \hat{A}^\dagger \hat{A} \rangle_{\rm final}  \right] e^{-
\gamma_{\rm eff} \tau},
\end{equation} as dictated by the quantum Langevin equation [Eq. (\ref{qle4})]. Thus referring to the quantum Otto cycle, during the isochoric steps $(b) \rightarrow (c)$ and $(d) \rightarrow (a)$, one can model the relaxation of the occupation number with respect to the photon and phonon baths as an exponential approach to the corresponding equilibrium values. Denoting $N_i = \langle \hat{A}^\dagger \hat{A} \rangle_i$ and $N_f = \langle \hat{A}^\dagger \hat{A} \rangle_f$, the relaxation dynamics on each isochore is given by
\begin{equation}
      N_c \approx N_f + \big[\langle \hat{A}^\dagger \hat{A} \rangle_b - N_f\big] e^{-\gamma_0 \tau_{bc}},
\end{equation}
\begin{equation}
 N_a \approx N_i + \big[\langle \hat{A}^\dagger \hat{A} \rangle_d - N_i\big] e^{-\gamma_m \tau_{da}},
\end{equation}
where $N_c = \langle \hat{A}^\dagger \hat{A} \rangle_{(b\rightarrow c)}(\tau_{bc})$ and $N_a = \langle \hat{A}^\dagger \hat{A} \rangle_{(d\rightarrow a)}(\tau_{da})$, with $\tau_{bc}$ and $\tau_{da}$ being the durations of the cold and hot isochores, respectively. In writing the above-mentioned expressions, we have approximately expressed $\gamma_{\rm eff} \approx \gamma_0$ for the step $(b) \rightarrow (c)$ and $\gamma_{\rm eff} \approx \gamma_m$ for the step $(d) \rightarrow (a)$. Using shortcuts to adiabaticity along the isentropes ensures that the populations are conserved during the strokes $(a)\rightarrow(b)$ and $(c)\rightarrow(d)$, so that $\langle \hat{A}^\dagger \hat{A} \rangle_b = N_a$ and $\langle \hat{A}^\dagger \hat{A} \rangle_d = N_c$. Introducing the factors $\zeta = e^{-\gamma_0 \tau_{bc}}$ and $ \xi = e^{-\gamma_m \tau_{da}}$, some algebra reveals the expressions
\begin{equation}
    N_c = N_f + (N_a - N_f)\zeta, \quad \quad
    N_a = N_i + (N_c - N_i)\xi,
\end{equation}
from which one obtains the population difference between the hot and cold ends of the cycle as given by
\begin{equation}
    N_a - N_c 
    = \frac{(1-\xi)(1-\zeta)}{1-\xi \zeta}(N_i - N_f).
\end{equation}
Since the work done per cycle and the heat absorbed from the hot bath are, respectively, given by the generic expressions
\begin{eqnarray}
    W^{\rm (finite~time)} 
    &=& \hbar(\Omega_i - \Omega_f)(N_a - N_c), \nonumber \\
    Q^{\rm (finite~time)}_{\rm in} 
    &=& \hbar\Omega_i (N_a - N_c),
\end{eqnarray}
one immediately observes that they are both reduced by the same multiplicative factor with respect to their quasistatic values $W^{\rm (ideal)} = \hbar(\Omega_i-\Omega_f)(N_i-N_f)$ and $Q^{\rm (ideal)}_{\rm in} = \hbar\Omega_i(N_i-N_f)$. Crucially, the efficiency
\begin{equation}
    \eta^{\rm (finite~time)} 
    = \frac{W^{\rm (finite~time)}}{Q^{\rm (finite~time)}_{\rm in}} 
    = 1 - \frac{\Omega_f}{\Omega_i}
\end{equation}
remains identical to the ideal efficiency [Eq. (\ref{otto_eff_general})]. That is, incomplete thermalization reduces both the work output and the absorbed heat by the same factor, leaving their ratio unchanged. This structure is fully consistent and relies on the linearized dynamics together with shortcuts to adiabaticity that suppress nonadiabatic excitations during the isentropic strokes.

\section{Conclusions}\label{conc_sec}
In this paper, we have put forward a theoretical framework for quantum heat engines powered by a ring-trapped BEC placed in a cavity by exploiting detuning sweeps of a polariton branch. To set the stage, we have diagonalized the linearized Hamiltonian, derived the Hopfield weights, and used quantum Langevin equations to bridge dissipation and fluctuations to thermodynamics. Our analysis leads to analytical expressions for work and efficiency along the lower polariton branch. Finite-time scenarios were also discussed which includes adopting shortcuts to adiabaticity. Provided both the isentropes are frictionless (via adiabatic shortcuts), the adiabatic strokes transport identical populations between the isochores; thus the efficiency remains the same, irrespective of incomplete thermalization. The parameters considered in this work suggest that the proposal may be compatible with current experimental capabilities, although its full realization will require further 
investigation. Several natural extensions remain open, including the role of measurement backaction and the exploration of nonequilibrium signatures in ring-trapped BECs beyond the optomechanical linearization studied here. Taken together, these directions indicate that cavity-assisted OAM control of ring-trapped BECs could, at least in principle, offer a route to tunable heat-engine behavior, motivating future experimental and theoretical studies.\\

\textbf{Acknowledgements:} We are grateful to the anonymous referees for several insightful comments and suggestions which have substantially improved the paper. We are also thankful to Himadri Shekhar Dhar, Rahul Gupta, and Harsh Sharma for their interest in this work. A.G. gratefully acknowledges stimulating discussions with Jasleen Kaur and Sushanta Dattagupta on quantum heat engines. M.B. thanks the Air Force Office of Scientific Research (AFOSR) (FA9550-23-1-0259) for support.

\begin{widetext}

\appendix 

\section{Bogoliubov-dressed modes and neglecting atomic interactions}\label{appA}
Employing the Bogoliubov theory of weakly-interacting Bose gases, one actually finds that $\omega_{c,d}$ are to be replaced by the Bogoliubov-dressed frequencies \cite{Kanamoto_2003}
\begin{equation}
    \omega_{c,d}' = \sqrt{\omega_{c,d} (\omega_{c,d} + 4\tilde{g}N)}.
\end{equation}
Choosing the parameters $a_{\rm Na} = 0.1$ nm, $\omega_\rho/2\pi = 840$ s$^{-1}$, $R = 10~\mu$m, $N = 10^4$, and $\gamma_0 = 2\pi \times 10^3$ s$^{-1}$ \cite{Ludlow_2007}, with $g=2\hbar\omega_{\rho}a_{\mathrm{Na}}/R$, one finds $4\tilde{g}N = gN/\pi \hbar \approx 0.05\gamma_0$, giving 
\begin{equation}
\frac{\omega'_{c,d} - \omega_{c,d}}{\omega_{c,d}} \sim 10^{-4},
\end{equation}
far smaller than the detuning-induced polaritonic shifts explored in the engine cycle. Thus $\omega_{c,d}' \approx \omega_{c,d}$ and $\omega_{c,d} + 4\tilde{g}N \approx \omega_{c,d}$. Moreover, the terms $2\tilde{g}N (\hat{c}\hat{d} + \hat{c}^\dagger \hat{d}^\dagger)$ can be dropped in the rotating-wave approximation due to their rapidly-oscillating behavior. These facts have allowed us to drop the interatomic interactions.

\section{Linearization and rotating-wave approximation}\label{appB}
Starting with the Hamiltonian quoted in Eq. (\ref{ring_BEC_Hamiltonian}), where $\hat{X}_{c(d)}=(\hat{c}_{(d)}+\hat{c}_{(d)}^\dagger)/\sqrt{2}$, let us displace each mode about its classical mean value according to
\begin{equation}
\hat{a}=\bar{a}+\hat{\tilde{a}},\quad \quad 
\hat{c}=\alpha_c+\hat{\tilde{c}},\quad \quad 
\hat{d}=\alpha_d+\hat{\tilde{d}},
\end{equation}
where the two-tone intracavity field is
\begin{equation}
\bar{a}(t)=\alpha_2+\alpha_1 e^{i\delta t},
\end{equation} with $\delta=\omega_{L2}-\omega_{L1}$ being the tone separation. Separating the mean and fluctuation quadratures as
$\hat{X}_{c(d)}=\bar{X}_{c(d)}+\hat{\tilde{X}}_{c(d)}$, with $\bar{X}_{c(d)}=(\alpha_{c(d)}+\alpha_{c(d)}^*)/\sqrt{2}$,
let us expand the Hamiltonian in powers of the fluctuation operators and retain terms up to second order. Constant terms are discarded and linear terms vanish when the mean amplitudes satisfy the classical equations of motion. The resulting Hamiltonian reads (after dropping the weak interatomic interactions)
\begin{equation}
\frac{\hat{H}_{\rm lin}}{\hbar}
=
-\bar{\Delta}\hat{\tilde{a}}^\dagger\hat{\tilde{a}}
+\omega_c\hat{\tilde{c}}^\dagger\hat{\tilde{c}}
+\omega_d\hat{\tilde{d}}^\dagger\hat{\tilde{d}}
+G(\hat{\tilde{X}}_c+\hat{\tilde{X}}_d)\Big[\bar{a}^*(t)\hat{\tilde{a}}+\bar{a}(t)\hat{\tilde{a}}^\dagger\Big],
\label{app_Hlin}
\end{equation}
where the effective detuning includes the static radiation-pressure-induced shift as
\begin{equation}
\bar{\Delta}=\tilde\Delta_2-G(\bar{X}_c+\bar{X}_d).
\end{equation}
Introducing the linearized coupling
$g(t)=G\bar{a}(t)/\sqrt{2}$, the interaction term can be written, for each mode
$j\in\{c,d\}$, as
\begin{equation}
\frac{\hat{H}_{\rm int}^{(j)}}{\hbar}
=
g(t)\hat{\tilde{a}}^\dagger \hat{\tilde{j}} + g^*(t)\hat{\tilde{a}}\hat{\tilde{j}}^\dagger
+
g(t)\hat{\tilde{a}}^\dagger \hat{\tilde{j}}^\dagger + g^*(t)\hat{\tilde{a}} \hat{\tilde{j}},
\label{app_BS_TMS}
\end{equation}
corresponding to beam-splitter and two-mode-squeezing interactions, respectively. To justify the rotating-wave approximation, let us move to the interaction picture defined by
\begin{equation}\label{H0}
\frac{\hat{H}_0}{\hbar}
=
-\bar{\Delta}\hat{\tilde{a}}^\dagger\hat{\tilde{a}}
+\omega_c\hat{\tilde{c}}^\dagger\hat{\tilde{c}}
+\omega_d\hat{\tilde{d}}^\dagger\hat{\tilde{d}}.
\end{equation}
In this frame, the beam-splitter terms oscillate at frequencies
$|\bar{\Delta}+\omega_j|$, while the two-mode-squeezing terms oscillate at
$-\bar{\Delta}+\omega_j$ ($j=c,d$). In the parameter regime considered here, $-\bar{\Delta}+\omega_j$ remains large throughout the detuning sweeps, satisfying
\begin{equation}
-\bar{\Delta}+\omega_{c,d} \gg \gamma_0,\ \gamma_m,\ |g(t)|,
\end{equation}
so that the squeezing terms are rapidly rotating and average out.
One thus obtains, after applying the rotating-wave approximation, the effective beam-splitter Hamiltonian
\begin{equation}
\frac{\hat{H}_{\rm eff}}{\hbar}
=
-\bar{\Delta}\hat{\tilde{a}}^\dagger\hat{\tilde{a}}
+\omega_c\hat{\tilde{c}}^\dagger\hat{\tilde{c}}
+\omega_d\hat{\tilde{d}}^\dagger\hat{\tilde{d}}
+\sum_{j=c,d}\Big[g_j\hat{\tilde{a}}^\dagger \hat{\tilde{j}} + g_j^*\hat{\tilde{a}}\hat{\tilde{j}}^\dagger\Big],
\label{app_Heff}
\end{equation}
where $g_c\propto G\alpha_1$ and $g_d\propto G\alpha_2$ are effective couplings determined by the intracavity amplitudes of the two tones. Although the linearized coupling
$g(t)=\frac{G}{\sqrt{2}}(\alpha_2+\alpha_1 e^{i\delta t})$
is explicitly time dependent in the rotating frame of the second control tone,
the effective couplings in Eq. (\ref{app_Heff}) are approximately stationary
once the rotating-wave approximation is performed.
To see this explicitly, let us write
\begin{equation}
g(t)=g_2+g_1 e^{i\delta t},
\quad \quad
g_{1,2}=\frac{G}{\sqrt{2}}\alpha_{1,2}.
\end{equation}
In the interaction picture generated by $H_0$ [Eq. (\ref{H0})],
the beam-splitter operator products evolve with oscillation frequencies $\bar{\Delta}+\omega_j$, i.e.,
\begin{equation}
\hat{\tilde{a}}^\dagger(t)\hat{\tilde{j}}(t)\propto
\hat{\tilde{a}}^\dagger \hat{\tilde{j}}e^{- i(\bar{\Delta}+\omega_j)t}.
\end{equation}
Substituting into the beam-splitter part of Eq. (\ref{app_BS_TMS}) yields the frequency component of each sidemode. In the rotating frame of tone 2, the $g_2$
contribution is resonant when $\bar{\Delta}+\omega_j\simeq 0$, whereas the $g_1$
contribution is resonant when $\bar{\Delta}+\omega_j-\delta\simeq 0$.
Choosing the tone separation $\delta$ to be $\delta\simeq \omega_c-\omega_d$, one can address the frequency difference between the sidemodes and excite them simultaneously. In particular, operating near $\bar{\Delta}\simeq -\omega_d$ makes
the $g_2$ term resonant for $j=d$ (since $\bar{\Delta}+\omega_d \simeq 0$), while
the $g_1$ term becomes simultaneously resonant for $j=c$ (since
$\bar{\Delta}+\omega_c-\delta \simeq -\omega_d+\omega_c-(\omega_c-\omega_d)=0$). In this configuration, the cross-coupling terms (tone 1 driving $d$ and tone 2
driving $c$) remain off-resonant by a large detuning $\sim|\omega_c-\omega_d| \gg
\gamma_0, \gamma_m, |g_{1,2}|$. Consequently, these rapidly-rotating cross terms
average out, and the remaining near-resonant contributions define stationary
effective couplings $g_c$ and $g_d$ in Eq. (\ref{app_Heff}) (up to fixed
phases that can be absorbed), set by the intracavity amplitudes of the two tones.

\vspace{2mm}

Sweeping the effective detuning $\bar{\Delta}$ as required for operating the heat engine therefore probes the system's
response as a function of proximity to the resonance conditions
$\bar{\Delta}+\omega_j\simeq 0$ and $\bar{\Delta}+\omega_j-\delta\simeq 0$,
without reintroducing explicit time dependence into the effective Hamiltonian.
By appropriately choosing the input drive strengths of the two tones, the couplings may be made
equal, i.e., $\tilde{G}=|g_c|=|g_d|$. Finally, relabeling $(\hat{\tilde{a}},\hat{\tilde{c}},\hat{\tilde{d}})=(\hat{a},\hat{c},\hat{d})$, one obtains the Hamiltonian
written in Eq. (\ref{Hfull}) of the main text.

\section{Normal-mode frequencies}\label{appC}
The characteristic polynomial found by setting $|\Lambda - \lambda I| = 0$ can be cast in the form
\begin{equation}\label{cubiceqnlambda}
    \lambda^3 + c_1\lambda^2 + c_2\lambda + c_3 = 0,
\end{equation}
where
\begin{equation}
    c_1 = \bar{\Delta} - \omega_c - \omega_d,  \quad \quad c_2 = \omega_c\omega_d - \bar{\Delta}(\omega_c+\omega_d) - 2\tilde{G}^2,  \quad \quad  c_3 = \tilde{G}^2(\omega_c+\omega_d) + \bar{\Delta}\omega_c\omega_d.
\end{equation}
Given a cubic equation of the form given in Eq. (\ref{cubiceqnlambda}) for some real-valued $(c_1,c_2,c_3)$, let us define the parameters
\begin{equation}
    p = c_2 - \frac{c_1^2}{3},\quad \quad
    q = \frac{2c_1^3}{27} - \frac{c_1 c_2}{3} + c_3.
\end{equation}
From the well-known Cardano's formula \cite{Bender_Orszag_1999}, the three real roots turn out to be
\begin{equation}
    R_k = -\frac{c_1}{3} + \chi\cos\left(\frac{\theta - 2\pi k}{3}\right),\quad \quad  \chi = 2\sqrt{-\frac{p}{3}}, \quad \quad \theta = \arccos\!\left(\frac{3q}{2p} \sqrt{-\frac{3}{p}}\right),     \end{equation}
where $k=0,1,2$.

\section{Polaritonic frequencies treating $\tilde{G}$ perturbatively}\label{appD}
Let us state here the behavior of the polaritonic modes for extreme limits of the detuning. This is achieved by treating $\tilde{G}$ perturbatively in the light of the Schrieffer-Wolff perturbation theory \cite{Bravyi_2011,Sakurai_Napolitano_2017}. The relevant regime for this expansion is
\begin{equation}
    \tilde{G} \ll |\omega_{c,d} + \bar{\Delta}|.
\end{equation}
The Schrieffer-Wolff perturbation theory begins with writing the Hamiltonian [Eq. (\ref{Hfull})] as
\begin{equation}
    \hat{H} = \hat{H}_0 + \hat{V},
\end{equation}
where
\begin{equation}
    \frac{\hat{H}_0}{\hbar} = -\bar{\Delta} \hat{a}^{\dagger} \hat{a} + \omega_{c} \hat{c}^{\dagger} \hat{c} + \omega_{d} \hat{d}^{\dagger} \hat{d} 
\end{equation}
is the non-interacting piece, while the effective light-matter interactions are given by
\begin{equation}
\frac{\hat{V}}{\hbar} = \tilde{G}(\hat{a}^{\dagger}\hat{c} + \hat{a}\hat{c}^{\dagger}) + \tilde{G}(\hat{a}^{\dagger}\hat{d} + \hat{a}\hat{d}^{\dagger}).
\end{equation}
In the regimes $-\bar{\Delta} \ll \omega_{c,d}$ or $-\bar{\Delta} \gg \omega_{c,d}$, one can define a Schrieffer-Wolff transformation based on an anti-Hermitian generator $\hat{S}$ such that the transformed Hamiltonian
\begin{equation}
    \hat{H}' = e^{\hat{S}} \hat{H} e^{-\hat{S}}
\end{equation}
is block‐diagonal up to the desired order in $\tilde{G}/\omega_{c,d}$. The defining condition $[\hat{H}_0, \hat{S}] = -\hat{V}$ eliminates the leading off-resonant couplings. Solving this for $\hat{S}$ yields
\begin{equation}
    \hat{S} \sim 
\frac{\tilde{G}}{\omega_{c}+\bar{\Delta}}(\hat{a}^{\dagger}\hat{c} - \hat{a}\hat{c}^{\dagger})
+
\frac{\tilde{G}}{\omega_{d}+\bar{\Delta}}(\hat{a}^{\dagger}\hat{d} - \hat{a}\hat{d}^{\dagger}),
\end{equation}
with the relative signs chosen so that the unwanted $\hat{a}\leftrightarrow \hat{c}$ and $\hat{a}\leftrightarrow \hat{d}$ couplings are canceled to the first order. Substituting this generator into
\begin{equation}
    \hat{H}' = \hat{H}_0 + \frac{1}{2}[\hat{V},\hat{S}] + \mathcal{O}(\tilde{G}^3)
\end{equation}
produces AC‐Stark shifts $\sim \tilde{G}^2/(\omega_{c,d}+\bar{\Delta})$. This directly allows one to determine the behavior of the polaritonic frequencies for large and small values of the detuning from the perturbative expansion described above. In particular, one obtains the following results with straightforward algebra.

\begin{enumerate}
    \item For $-\bar{\Delta} \ll \omega_{c,d}$:
    \begin{eqnarray}
        \omega_A &\simeq& -\bar{\Delta}-\frac{\tilde{G}^2}{\bar{\Delta}+\omega_c}-\frac{\tilde{G}^2}{\bar{\Delta}+\omega_d}, \label{omegaA11}\\
        \omega_B &\simeq& \omega_c+\frac{\tilde{G}^2}{\bar{\Delta}+\omega_c}, \\
        \omega_C &\simeq& \omega_d+\frac{\tilde{G}^2}{\bar{\Delta}+\omega_d}.
    \end{eqnarray}
\item For $-\bar{\Delta} \gg \omega_{c,d}$:
\begin{eqnarray}
        \omega_A &\simeq& \omega_d+\frac{\tilde{G}^2}{\bar{\Delta}+\omega_d}, \label{omegaA22} \\
        \omega_B &\simeq& -\bar{\Delta} -\frac{\tilde{G}^2}{\bar{\Delta} +\omega_c}-\frac{\tilde{G}^2}{\bar{\Delta} +\omega_d}, \\
        \omega_C &\simeq& \omega_c+\frac{\tilde{G}^2}{\bar{\Delta}+\omega_c}.
    \end{eqnarray}
\end{enumerate}
Thus switching $-\bar{\Delta}$ between the two regimes, one can make `$A$' switch between the photonlike and phononlike regimes and similarly one can switch `$B$' between the phononlike and photonlike regimes. 

\end{widetext}


\begin{thebibliography}{99}

\bibitem{Kosloff_2014}
R. Kosloff and A. Levy, {\it Quantum heat engines and refrigerators: Continuous devices}, Annu. Rev. Phys. Chem. \textbf{65}, 365 (2014). 

\bibitem{Cangemi_2024}
L. M. Cangemi, C. Bhadra, and A. Levy, {\it Quantum engines and refrigerators}, Phys. Rep. \textbf{1087}, 1 (2024). 

\bibitem{Scovil_1959}
H. E. D. Scovil and E. O. Schulz-DuBois, {\it Three-level masers as heat engines}, Phys. Rev. Lett. \textbf{2}, 262 (1959).

\bibitem{Alicki_1979}
R. Alicki, {\it The quantum open system as a model of a heat engine}, J. Phys. A: Math. Gen. \textbf{12}, L103 (1979).

\bibitem{Cavina_2017}
V. Cavina, A. Mari, and V. Giovannetti, {\it Slow dynamics and thermodynamics of open quantum systems}, Phys. Rev. Lett. \textbf{119}, 050601 (2017).

\bibitem{Scandi_2019}
M. Scandi and M. Perarnau-Llobet, {\it Thermodynamic length in open quantum systems}, Quantum \textbf{3}, 197 (2019).

\bibitem{Feldmann_2000}
T. Feldmann and R. Kosloff, {\it Performance of discrete heat engines and heat pumps in finite time}, Phys. Rev. E \textbf{61}, 4774 (2000).

\bibitem{Rezek_2006}
Y. Rezek and R. Kosloff, {\it Irreversible performance of a quantum harmonic heat engine}, New J. Phys. \textbf{8}, 83 (2006).

\bibitem{Chen_2010}
X. Chen, A. Ruschhaupt, S. Schmidt, A. del Campo, D. Gu\'ery-Odelin, and J. G. Muga, {\it Fast optimal frictionless atom cooling in harmonic traps: Shortcut to adiabaticity}, Phys. Rev. Lett. \textbf{104}, 063002 (2010).

\bibitem{delCampo_2013}
A. del Campo, {\it Shortcuts to adiabaticity by counterdiabatic driving}, Phys. Rev. Lett. \textbf{111}, 100502 (2013).

\bibitem{delCampo_2014}
A. del Campo, J. Goold, and M. Paternostro, {\it More bang for your buck: Super-adiabatic quantum engines}, Sci. Rep. \textbf{4}, 6208 (2014).

\bibitem{Quan_2007}
H. T. Quan, Y.-x. Liu, C. P. Sun, and F. Nori, {\it Quantum thermodynamic cycles and quantum heat engines}, Phys. Rev. E \textbf{76}, 031105 (2007).

\bibitem{Scully_2001}
M. O. Scully, {\it Extracting work from a single thermal bath via quantum negentropy}, Phys. Rev. Lett. \textbf{87}, 220601 (2001).

\bibitem{Kosloff_2013}
R. Kosloff, {\it Quantum thermodynamics: A dynamical viewpoint}, Entropy \textbf{15}, 2100 (2013).

\bibitem{Guarnieri_2019}
G. Guarnieri, G. T. Landi, S. R. Clark, and J. Goold, {\it Thermodynamics of precision in quantum nonequilibrium steady states}, Phys. Rev. Research \textbf{1}, 033021 (2019).

\bibitem{Bedkihal_2025}
S. Bedkihal, J. Behera, and M. Bandyopadhyay, {\it Fundamental aspects of Aharonov-Bohm quantum machines: Thermoelectric heat engines and diodes}, J. Phys.: Condens. Matter \textbf{37}, 163001 (2025).

\bibitem{Abah_2012}
O. Abah, J. Ro\textbeta nagel, G. Jacob, S. Deffner, F. Schmidt-Kaler, K. Singer, and E. Lutz, {\it Single-ion heat engine at maximum power}, Phys. Rev. Lett. \textbf{109}, 203006 (2012).

\bibitem{Rossnagel_2016}
J. Ro\textbeta nagel, S. T. Dawkins, K. N. Tolazzi, O. Abah, E. Lutz, F. Schmidt-Kaler, and K. Singer, {\it A single-atom heat engine}, Science \textbf{352}, 325 (2016).

\bibitem{Maslennikov_2019}
G. Maslennikov, S. Ding, R. Habl\"utzel, J. Gan, A. Roulet, S. Nimmrichter, J. Dai, V. Scarani, and D. Matsukevich, {\it Quantum absorption refrigerator with trapped ions}, Nat. Commun. \textbf{10}, 202 (2019).

\bibitem{Kosloff_Rezek_2017}
R. Kosloff and Y. Rezek, {\it The quantum harmonic Otto cycle}, Entropy \textbf{19}, 136 (2017).

\bibitem{Zhang_2014}
K. Zhang, F. Bariani, and P. Meystre, {\it Quantum optomechanical heat engine}, Phys. Rev. Lett. \textbf{112}, 150602 (2014).

\bibitem{Ivander_2022}
F. Ivander, N. Anto-Sztrikacs, and D. Segal, {\it Strong system-bath coupling effects in quantum absorption refrigerators}, Phys. Rev. E \textbf{105}, 034112 (2022).

\bibitem{Gelbwaser_2015}
A. Gelbwaser-Klimovsky, W. Niedenzu, and G. Kurizki, {\it Thermodynamics of quantum systems under dynamical control}, Adv. At. Mol. Opt. Phys. \textbf{64}, 329 (2015).

\bibitem{Brantut_2013}
J.-P. Brantut, C. Grenier, J. Meineke, D. Stadler, S. Krinner, C. Kollath, T. Esslinger, and A. Georges, {\it A thermoelectric heat engine with ultracold atoms}, Science \textbf{342}, 713 (2013).

\bibitem{Barontini_2019}
G. Barontini and M. Paternostro, {\it Ultra-cold single-atom quantum heat engines}, New J. Phys. \textbf{21}, 063019 (2019).

\bibitem{Koch_2023}
J. Koch, K. Menon, E. Cuestas, S. Barbosa, E. Lutz, T. Fogarty, T. Busch, and A. Widera, {\it A quantum engine in the BEC-BCS crossover}, Nature \textbf{621}, 723 (2023).

\bibitem{Simmons_2023}
E. Q. Simmons, R. Sajjad, K. Keithley, H. Mas, J. L. Tanlimco, E. Nolasco-Martinez, Y. Bai, G. H. Fredrickson, and D. M. Weld, {\it Thermodynamic engine with a quantum degenerate working fluid}, Phys. Rev. Research \textbf{5}, L042009 (2023).

\bibitem{Ruan_2024}
H. Ruan, J. Yuan, Y. Xu, J. He, Y. Ma, and J. Wang, {\it Performance enhancement of quantum Brayton engine via Bose-Einstein condensation}, Phys. Rev. E \textbf{109}, 024126 (2024).

\bibitem{Morizot_2006}
O. Morizot, Y. Colombe, V. Lorent, H. Perrin, and B. M. Garraway, {\it Ring trap for ultracold atoms}, Phys. Rev. A \textbf{74}, 023617 (2006). 

\bibitem{Wright_2013}
K. C. Wright, R. B. Blakestad, C. J. Lobb, W. D. Phillips, and G. K. Campbell, {\it Driving phase slips in a superfluid atom circuit with a rotating weak link}, Phys. Rev. Lett. \textbf{110}, 025302 (2013).

\bibitem{Molina-Terriza_2001}
G. Molina-Terriza, J. P. Torres, and L. Torner, {\it Management of the angular momentum of light: Preparation of photons
in multidimensional vector states of angular momentum}, Phys. Rev. Lett. \textbf{88}, 013601 (2001).

\bibitem{Yao_2011}
A. M. Yao and M. J. Padgett, {\it Orbital angular momentum: Origins, behavior and applications}, Adv. Opt. Photon. \textbf{3}, 161 (2011).

\bibitem{Fickler_2012}
R. Fickler, R. Lapkiewicz, W. N. Plick, M. Krenn, C. Schaeff, S. Ramelow, and A. Zeilinger, {\it Quantum entanglement of high angular momenta}, Science \textbf{338}, 640 (2012).

\bibitem{Brennecke_2008}
F. Brennecke, S. Ritter, T. Donner, and T. Esslinger, {\it Cavity optomechanics with a Bose-Einstein condensate}, Science \textbf{322}, 235 (2008).

\bibitem{Pandey_2019}
S. Pandey, H. Mas, G. Drougakis, P. Thekkeppatt, V. Bolpasi, G. Vasilakis, K. Poulios, and W. von Klitzing, {\it Hypersonic Bose-Einstein condensates in accelerator rings}, Nature \textbf{570}, 205 (2019).

\bibitem{Kumar_2021}
P. Kumar, T. Biswas, K. Feliz, R. Kanamoto, M.-S. Chang, A. K. Jha, and M. Bhattacharya, {\it Cavity optomechanical sensing and manipulation of an atomic persistent current}, Phys. Rev. Lett. \textbf{127}, 113601 (2021).

\bibitem{Kalita_2023}
S. Kalita, P. Kumar, R. Kanamoto, M. Bhattacharya, and A. K. Sarma, {\it Pump-probe cavity optomechanics with a rotating atomic superfluid in a ring}, Phys. Rev. A \textbf{107}, 013525 (2023).

\bibitem{1_Pradhan_2024}
N. Pradhan, P. Kumar, R. Kanamoto, T. N. Dey, M. Bhattacharya, and P. K. Mishra, {\it Cavity optomechanical detection of persistent currents and solitons in a bosonic ring condensate}, Phys. Rev. Research \textbf{6}, 013104 (2024).

\bibitem{2_Pradhan_2024}
N. Pradhan, P. Kumar, R. Kanamoto, T. N. Dey, M. Bhattacharya, and P. K. Mishra, {\it Ring Bose-Einstein condensate in a cavity: Chirality detection and rotation sensing}, Phys. Rev. A \textbf{109}, 023524 (2024).

\bibitem{Das_2024}
S. Das, P. Kumar, M. Bhattacharya, and T. N. Dey, {\it Hybrid rotational cavity optomechanics using an atomic superfluid in a ring}, Phys. Rev. A \textbf{110}, 043512 (2024).

\bibitem{Gupta_2024}
R. Gupta, P. Kumar, R. Kanamoto, M. Bhattacharya, and H. S. Dhar, {\it Sensing atomic superfluid rotation beyond the standard quantum limit}, Phys. Rev. A \textbf{110}, 053514 (2024).

\bibitem{Pradhan_2025}
N. Pradhan, R. Kanamoto. M. Bhattacharya, and P. K. Mishra, {\it Signature of Andreev-Bashkin superfluid drag from cavity optomechanics}, Phys. Rev. Research \textbf{7}, 023051 (2025).

\bibitem{Gerry_Knight_2004}
C. Gerry and P. Knight, {\it Beam splitters and interferometers}, In: Introductory quantum optics, Cambridge University Press (2004).

\bibitem{Aspelmeyer_2014}
M. Aspelmeyer, T. J. Kippenberg, and F. Marquardt, {\it Cavity optomechanics}, Rev. Mod. Phys. \textbf{86}, 1391 (2014).

\bibitem{Ludlow_2007}
A. D. Ludlow, X. Huang, M. Notcutt, T. Zanon-Willette, S. M. Foreman, M. M. Boyd, S. Blatt, and J. Ye, {\it Compact, thermal-noise-limited optical cavity for diode laser stabilization at $1 \times 10^{-15}$}, Opt. Lett. \textbf{32}, 641 (2007).

\bibitem{Hopfield_1958}
J. J. Hopfield, {\it Theory of the contribution of excitons to the complex dielectric constant of crystals}, Phys. Rev. \textbf{112}, 1555 (1958).

\bibitem{Kavokin_Malpuech_2003}
A. Kavokin and G. Malpuech, {\it Cavity polaritons}, Elsevier/Academic Press (2003).

\bibitem{Gardiner_Zoller_2004}
C. Gardiner and P. Zoller, {\it Quantum noise: A handbook of Markovian and non-Markovian quantum stochastic methods with applications to quantum optics}, Springer (2004).

\bibitem{Seifert_2012}
U. Seifert, {\it Stochastic thermodynamics, fluctuation theorems and molecular machines}, Rep. Prog. Phys. \textbf{75}, 126001 (2012). 

\bibitem{Moulder_2012}
S. Moulder, S. Beattie, R. P. Smith, N. Tammuz, and Z. Hadzibabic, {\it Quantized supercurrent decay in an annular Bose-Einstein condensate}, Phys. Rev. A \textbf{86}, 013629 (2012).

\bibitem{Pethick_Smith_2001}
C. J. Pethick and H. Smith, {\it Bose-Einstein condensation in dilute gases}, Cambridge University Press (2001).

\bibitem{Breuer_Petruccione_2002}
H.-P. Breuer and F. Petruccione, {\it The theory of open quantum systems}, Oxford University Press (2002). 

\bibitem{Husimi_1953}
K. Husimi, {\it Miscellanea in elementary quantum mechanics, II}, Prog. Theor. Phys. \textbf{9}, 381 (1953).

\bibitem{Kiran_2021}
T. Kiran and M. Ponmurugan, {\it Invariant-based investigation of shortcut to adiabaticity for quantum harmonic oscillators under a time-varying frictional force}, Phys. Rev. A \textbf{103}, 042206 (2021).

\bibitem{Morris_2015}
R. M. Morris and P. G. L. Leach, {\it The Ermakov-Pinney equation: Its varied origins and the effects of the introduction of symmetry-breaking functions}, arXiv:1510.08992.

\bibitem{Kanamoto_2003}
R. Kanamoto, H. Saito, and M. Ueda, {\it Quantum phase transition in one-dimensional Bose-Einstein condensates with attractive interactions}, Phys. Rev. A \textbf{67}, 013608 (2003).

\bibitem{Bender_Orszag_1999}
C. M. Bender and S. A. Orszag, {\it Advanced mathematical methods for scientists and engineers I: Asymptotic methods and perturbation theory}, Springer (1999).

\bibitem{Bravyi_2011}
S. Bravyi, D. P. DiVincenzo, and D. Loss,
{\it Schrieffer-Wolff transformation for quantum many-body systems},
Ann. Phys. \textbf{326}, 2793 (2011).

\bibitem{Sakurai_Napolitano_2017}
J. J. Sakurai and J. Napolitano, {\it Modern quantum mechanics}, 2nd ed., Cambridge University Press (2017).

\end{thebibliography}
\end{document}